\documentclass[aps,prb,twocolumn,showpacs]{revtex4}

\usepackage{hyperref,latexsym}
\usepackage{graphicx,color,pstricks}
\usepackage{epsfig,epsf,bm}

\begin{document}
\title{Quantum phase transition in an atomic Bose gas near a Feshbach resonance}
\author{Yu-Wen Lee}
\affiliation{Physics Department, Tunghai University, Taichung,
Taiwan, R.O.C.} \email{ywlee@mail.thu.edu.tw}
\author{Yu-Li Lee}
\affiliation{Physics Department, National Changhua University of
Education, Changhua, Taiwan, R.O.C.} \email{yllee@cc.ncue.edu.tw}
\begin{abstract}
 We study the quantum phase transition in an atomic Bose gas near
 a Feshbach resonance in terms of the renormalization group. This
 quantum phase transition is characterized by an Ising order
 parameter. We show that in the low temperature regime where the
 quantum fluctuations dominate the low-energy physics this phase
 transition is of first order because of the coupling between the
 Ising order parameter and the Goldstone mode existing in the bosonic
 superfluid. However, when the thermal fluctuations become
 important, the phase transition turns into the second order one,
 which belongs to the three-dimensional Ising universality class.
 We also calculate the damping rate of the collective mode in the
 phase with only a molecular Bose-Einstein condensate near the
 second-order transition line, which can serve as an experimental
 signature of the second-order transition.
\end{abstract}
 \maketitle

\section{Introduction}

Trapped dilute cold atomic gases are one of the most exciting
fields in condensed matter physics.\cite{A,D} An important recent
development in this area is the application of Feshbach
resonances. A Feshbach resonance in the scattering amplitude of
two atoms occurs when the total energy of the atoms is close to
the energy of a molecular state that is weakly coupled to the
atomic continuum. Especially, the energy difference between the
molecular state and the two-atom continuum, known as the detuning
$\delta$, can be experimentally tuned by means of a magnetic
field. Therefore, by sweeping the magnetic field from positive to
negative detuning through the Feshbach resonance, it is actually
possible to form molecules in the atomic gas.\cite{RTB,SPH,XMA} In
fact, recently, it has been possible to create a Bose-Einstein
condensate (BEC) of molecules in an atomic Fermi gas with a
Feshbach resonance.\cite{JBA,GRJ,ZSS} This offers the opportunity
for observing a Bardeen-Cooper-Schrieffer (BCS) transition in a
dilute Fermi gas.

In the present paper, we study an analogous situation by varying
$\delta$ in an atomic Bose gas with a Feshbach resonance. Two
recent works\cite{RPW,RDS} have shown that by varying $\delta$
there is a true quantum phase transition (QPT) in an atomic Bose
gas, in contrast to the case of an atomic Fermi gas where a smooth
BEC-BCS crossover exists as one changes $\delta$. An argument to
understand this has been given in Ref. \onlinecite{RDS}. We
recapitulate it in the following to fix our notation. The
effective Lagrangian describing a dilute atomic gas with a
Feshbach resonance can be written as
\begin{eqnarray}
 L &=& \Psi_a^{\dagger} \! \left(\partial_{\tau}-\frac{\nabla^2}
   {2m}-\mu\right) \! \Psi_a+\Psi_m^{\dagger} \! \left(
   \partial_{\tau}-\frac{\nabla^2}{2M}-\tilde{\mu}\right) \!
   \Psi_m \nonumber \\
   & & +g_0\left(\Psi^{\dagger}_m\Psi_a\Psi_a+{\mathrm H.c.}\right)
   +u_3\Psi^{\dagger}_m\Psi_a^{\dagger}\Psi_a\Psi_m \nonumber \\
   & & +\frac{u_1}{2}\Psi^{\dagger}_a\Psi_a^{\dagger}\Psi_a\Psi_a
   +\frac{u_2}{2}\Psi^{\dagger}_m\Psi_m^{\dagger}\Psi_m\Psi_m \ ,
   \label{feshbact1}
\end{eqnarray}
with $\tilde{\mu}=2\mu-\delta$. Here $\Psi_a$ and $\Psi_m$ are the
annihilation operators of atoms and molecules, respectively. $m$
and $M$ are the mass of the atom and that of the molecule,
respectively. In the path integral formula, $\Psi_m$ is an
ordinary number. On the other hand, $\Psi_a$ is an ordinary number
for the atomic Bose gas, while it is a Grassmann number for the
atomic Fermi gas. (For the atomic Fermi gas, $\Psi_a$ contains the
indices for hyperfine spins and should be understood as a spinor.)
The Lagrangian $L$ [Eq. (\ref{feshbact1})] has a U($1$) symmetry.
That is, it is invariant against the U($1$) transformation
\begin{equation}
 \Psi_a\rightarrow e^{i\alpha}\Psi_a \ , ~~
 \Psi_m\rightarrow e^{2i\alpha}\Phi_m \ . \label{bu11}
\end{equation}
For an atomic Bose gas, due to the $g_0$ term, a nonzero value of
$\langle \Psi_a\rangle$ (atomic BEC) must lead to a nonzero value
of $\langle \Psi_m\rangle$ (molecular BEC). However, the reverse
is not true. That is, it is possible for the gas to contain only a
molecular BEC. In the case with both atomic BEC and molecular BEC,
the U($1$) symmetry of $L$ is completely broken. But for the case
with only molecular BEC there is a residual Z$_2$ symmetry. That
is, the effective Lagrangian in this case is invariant against the
Z$_2$ transformation: $\Psi_a\rightarrow -\Psi_a$ and
$\Psi_m\rightarrow \Psi_m$. The above analysis indicates that at
low temperature the atomic Bose gas with a Feshbach resonance has
two thermodynamically distinct phases: the ``atomic superfluid"
(ASF) phase with both atomic BEC and molecular BEC, and the
``molecular superfluid" (MSF) phase with molecular BEC only. For
an atomic Fermi gas, due to the $g_0$ term, a nonzero value of
$\langle \Psi_a\Psi_a\rangle$ must be accompanied with a
nonvanishing value of $\langle \Psi_m\rangle$ and vice versa.
Therefore, the BCS region has the same symmetry as the BEC region,
and only a crossover occurs.

\begin{figure}
\begin{center}
 \includegraphics[width=0.8\columnwidth]{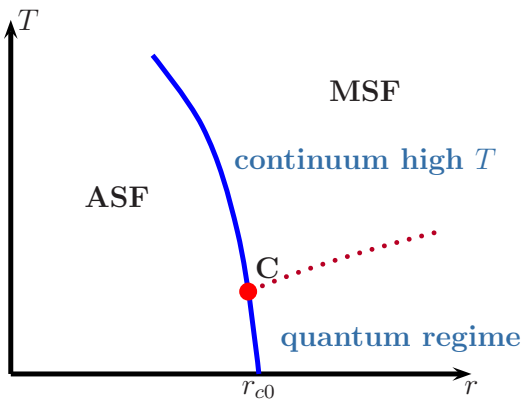}
 \caption{The schematic phase diagram of the dilute Bose gas with
         single Feshbach resonance near the transition line between
         the ASF and MSF phases below $T_c(\delta)$, where $T_c(\delta)$
         represents the transition temperature between the normal Bose
         gas and the superfluid phase. The dotted line denotes the
         crossover from low temperature to high temperature regimes.
         The solid line is the phase boundary between the ASF and MSF
         phases. The point $C$ is a tricritical point. On the phase
         boundary, the portion below the point $C$ is of first order
         while the one above the point $C$ is of second order.}
 \label{phase}
\end{center}
\end{figure}

Based on the above observation, the MSF phase is distinct from the
ASF phase by a Z$_2$ symmetry. Thus, the phase transition between
the two phases is characterized by an Ising (Z$_2$) order
parameter. Naively, one may expect that this QPT is of second
order and belongs to the four-dimensional ($4d$) Ising
universality class at zero temperature and the three-dimensional
($3d$) Ising universality class at finite temperature. However,
there is a gapless excitation in the MSF phase, which is the
Goldstone mode associated with the molecular BEC. An additional
gapless excitation may result in severe IR divergences and modify
the critical behavior. Therefore, a proper theory describing the
QPT should consist of the Ising order parameter as well as the
Goldstone mode. A renormalization group (RG) analysis indicates
that the coupling between the Ising order parameter and the
Goldstone mode will drive the zero-temperature phase transition to
become weakly first order\cite{FB} through the Colemann-Weinberg
mechanism.\cite{CW} The question in which we are interested here
is the nature of this phase transition at finite temperature. We
study this problem in terms of the RG. Our main results are shown
in Fig. \ref{phase}, which are valid at the temperature much lower
than the amplitude fluctuations in the molecular BEC. At low
temperature where the quantum fluctuations dominate the low-energy
physics, the phase transition between the MSF and ASF phases is of
weakly first order. When the thermal fluctuations become
important, the effects of the coupling between the Ising order
parameter and the Goldstone mode are suppressed and the transition
becomes a second-order one belonging to the $3d$ Ising
universality class. Thus, a tricritical point must exist on the
phase boundary to separate these two kinds of phase transition.

The rest of the paper is organized as follows: We write down the
effective Lagrangian describing the QPT by symmetry argument and
perform one-loop RG analysis in Sec. \ref{rg}. The solution of the
scaling equations is presented in Sec. \ref{sol}. In Sec.
\ref{damp}, we calculate the damping rate of the collective
excitation in the MSF phase near the second-order transition line.
The last section is devoted to our conclusion.

\section{Landau theory and renormalization group analysis}
\label{rg}
\subsection{Effective Lagrangian}

We start with the ``normal phase" for this QPT. The fundamental
fields of the effective theory describing the QPT consist of the
Ising order parameter $\phi$, which can be taken as the imaginary
part of $\Psi_a$, and the phase of $\Psi_m$, $\theta$. Following
Landau, the corresponding effective Lagrangian can be written down
through symmetry consideration. The symmetries involved here are
the Z$_2$ symmetry, a subgroup of the U($1$) symmetry [Eq.
(\ref{bu11})], under which the $\phi$ and $\theta$ fields
transform as
\begin{equation}
 \phi \rightarrow -\phi \ , ~~
 \theta \rightarrow \theta +2\pi \ , \label{z22}
\end{equation}
and the U($1$) symmetry under which $\theta$ transforms as $\theta
(x) \rightarrow \theta (x)+\alpha$ where $\alpha$ is an arbitrary
constant. The most general local action consistent with the Z$_2$
and U($1$) symmetries is $I=I_{\phi}+I_{\theta}+I_{int}$ where
\begin{equation}
 I_{\phi}=\! \! \int \! \! d\tau d^3x\left\{\frac{1}{2}\left[
         \left(\frac{1}{c_0}\partial_{\tau}\phi\right)^2 \! \!
         +(\bm{\nabla}\phi)^2\right] \! \! +\frac{t_0}{2}\phi^2
         +\frac{u_0}{4!}\phi^4 \! \right\} , \label{qcplag1}
\end{equation}
with $u_0>0$,
\begin{equation}
 I_{\theta}=\frac{\rho_s}{2M}\int \! \! d\tau d^3x\left[ \! \left(
           \frac{1}{v_0}\partial_{\tau}\theta \right)^2
           +(\bm{\nabla}\theta)^2\right] , \label{qcplag2}
\end{equation}
and
\begin{equation}
 I_{int}=\int \! \! d\tau d^3x~i\lambda_0\partial_{\tau}\theta
         \phi^2 \ . ~~~~~~~\label{qcplag3}
\end{equation}
Here $t_0$ measures the (mean-field) distance from the transition
point. ($t_0>0$ in the MSF phase while $t_0<0$ in the ASF phase.)
$\rho_s$ is the "bare" superfluid density and $c_0$ and $v_0$ are
"bare" velocities for the order parameter $\phi$ and the Goldstone
boson $\theta$, respectively. In Eq. (\ref{qcplag3}), only the
most relevant (near the transition point) coupling between the
Goldstone mode and the order parameter is kept. The factor $i$ in
Eq. (\ref{qcplag3}) is dictated by the charge conjugation symmetry
of the original Lagrangian $L$ [Eq. (\ref{feshbact1})]. It
requires that $L\rightarrow L^{\dagger}$ when $\Psi_a\rightarrow
\Psi_a^{\dagger}$ and $\Psi_m\rightarrow \Psi_m^{\dagger}$, which
leads to $I\rightarrow I^{\dagger}$ when $\phi \rightarrow -\phi$
and $\theta \rightarrow -\theta$. The natural cutoff of this
action is provided by the gap of amplitude fluctuations in the
molecular BEC. A similar action has appeared in other
context.\cite{FB}

The action (\ref{qcplag1}) --- (\ref{qcplag3}) can also be derived
from the Lagrangian $L$ [Eq. (\ref{feshbact1})] by considering the
fluctuations around its mean-field solution in the MSF phase. By
integrating out the sectors with finite gaps at the transition
point, i.e. the real part of $\Psi_a$ and the amplitude of
$\Psi_m$, in terms of the perturbation theory in $u_1$, $u_2$,
$u_3$, and $g_0/\sqrt{n}$, to the leading order, one may obtain
\begin{eqnarray}
 & & c_0\approx \sqrt{\frac{g_0}{m}}\left(2n\right)^{1/4} \ ,
     \nonumber \\
 & & t_0=m(\delta_0 -\delta) \ , \nonumber \\
 & & u_0\approx 12mu_1 \! \left[1-\frac{1}{u_1u_2}\left(u_3
     +\sqrt{\frac{2}{n}}~g_0\right)^2\right] , \nonumber \\
 & & \rho_s\approx \frac{n}{2} \ , \nonumber \\
 & & v_0\approx \sqrt{\frac{u_2n}{2M}} \ , \nonumber \\
 & & \lambda_0\approx \frac{m}{2}\left[1-\frac{2}{u_2}\left(u_3
     +\sqrt{\frac{2}{n}}~g_0\right)\right] , \label{lagc2}
\end{eqnarray}
near the transition point. (Here we take the real part of $\Psi_a$
as the gapped sector at the transition point. This amounts to
assuming $g_0>0$.) In Eq. (\ref{lagc2}), $n$ is the total particle
density and $\delta_0$ denotes the magnetic field detuning at
which the zero-temperature quantum phase transition occurs at the
mean-field level, which is given by $\delta_0\approx
\left(u_3-\frac{u_2}{2}\right)n-2^{3/2}g_0n^{1/2}$. (This value of
$\delta_0$ has already been given by Ref. \onlinecite{RPW}.) From
Eq. (\ref{lagc2}), it is obvious that the value of $u_0$ may
become negative in some parameter regime. For $u_0<0$, one should
include the terms with higher powers of $\phi$ like $\phi^6$ in
$I_{\phi}$ to make it stable, and the QPT between the MSF and ASF
phases will be of first order,\cite{KH} even in the absence of the
coupling to the Goldstone mode $\theta$. In this case, it is of
little interest because the correlation length is always finite
and thus there are no universal behaviors for physical quantities
at both zero and finite temperatures. Therefore, in the following,
we shall concentrated on the parameter regime which gives rise to
a positive value of $u_0$. From Eq. (\ref{lagc2}), this
corresponds to $(u_3+\sqrt{2/n}~g_0)^2<u_1u_2$.

At $T=0$, $I_{\phi}$ with $u_0>0$ describes the $4d$ Ising
universality class. It exhibits a Gaussian behavior with
logarithmic corrections due to the presence of the marginally
irrelevant coupling $u_0$. Therefore, we may consider the RG
transformation
\begin{eqnarray}
 & & \tau \rightarrow \tau /s \ , ~~ \bm{x}\rightarrow \bm{x}/s
     \ , \nonumber \\
 & & \phi \rightarrow s\phi \ , ~~ \theta \rightarrow s\theta
     \ , \label{qcprg1}
\end{eqnarray}
while $c_0$, $v_0$, $\rho_s$, and $M$ remain invariant. Here $s>1$
is the rescaling factor. Then, the action
$I_{\phi}+I_{\theta}+I_{int}$ with $u_0=0=\lambda_0$ is invariant
under the RG transformation [Eq. (\ref{qcprg1})]. It is
straightforward to see that the coupling constants $u_0$ and
$\lambda_0$ are marginal at the tree level with respect to the
Gaussian fixed point under the RG transformation [Eq.
(\ref{qcprg1})]. Within the weak-coupling region, one-loop RG
equations are needed to determine their roles on the low-energy
physics.

\subsection{One-loop RG equations}

To compute the one-loop RG equations, we employ the momentum-shell
RG. Before integrating out the fast modes, we make a change on the
variables:
\begin{eqnarray*}
 & & \bm{x}\rightarrow \Lambda^{-1}\bm{x} \ , ~~\tau \rightarrow
     \Lambda^{-1}c_0^{-1}\tau \ , ~~ v=v_0/c_0 \ , \\
 & & \phi \rightarrow \Lambda c_0^{1/2}\phi \ , ~~ \theta
     \rightarrow \Lambda \! \left(\frac{c_0M}{\rho_s}\right)^{1/2}
     \! \theta \ , \\
 & & r=\Lambda^{-2}t_0 \ , ~~u=c_0u_0 \ , ~~ \beta =\frac{\Lambda c_0}
     {T} \ , \\
 & & \lambda=\left(\frac{c_0^3M}{\rho_s}\right)^{1/2} \! \lambda_0
     \ ,
\end{eqnarray*}
where $\Lambda$ is an UV cutoff in momenta and $T$ is the
temperature. Then, $v$, $r$, $u$, $\lambda$, and $\beta$ become
dimensionless parameters and our working action is written as
$I=\int^{\beta}_0 \! \! d\tau \! \int \! \! d^3x~{\mathcal L}$
where
\begin{eqnarray}
 {\mathcal L} &=& \frac{1}{2}\left[(\partial_{\tau}\phi)^2+(\bm{\nabla}
              \phi)^2\right]+\frac{1}{2}~r\phi^2+\frac{1}{4!}~u\phi^4
              \nonumber \\
              & & +\frac{1}{2}\left[\left(\frac{1}{v}~\partial_{\tau}
              \theta\right)^2+(\bm{\nabla}\theta)^2\right]+i\lambda
              \partial_{\tau}\theta\phi^2 \ . ~~\label{wlag1}
\end{eqnarray}

By integrating out the fast modes, i.e. those modes with momenta
within the momentum shell $e^{-l}<|\bm{k}|<1$ where $e^{-l}$ is
the scaling factor, we obtain the one-loop RG equations:
\begin{eqnarray}
 & & \frac{dt_l}{dl}=t_l \ , \label{bfrge1} \\
 & & \frac{dv_l}{dl}=-\frac{K_lv_l}{4(1+r_l)^{3/2}}~
     f_2(\sqrt{1+r_l}/t_l) \ , \label{bfrge5} \\
 & & \frac{dr_l}{dl}=2r_l+\frac{U_lt_l}{4(1+r_l)}f_1(\sqrt{1+r_l}/t_l)
     \nonumber \\
 & & ~~~~~~~~+\frac{2K_l}{t_l}~g_1(\sqrt{1+r_l}/t_l,v_l/t_l) \ ,
     \label{bfrge2} \\
 & & \frac{dK_l}{dl}=-\frac{K_l}{4}(U_l+2K_l)\frac{f_2(\sqrt{1+r_l}/t_l)}
     {(1+r_l)^{3/2}} \nonumber \\
 & & ~~~~~~~~-\frac{2K_l^2}{\sqrt{1+r_l}(1+r_l-v_l^2)}~g_2
     (\sqrt{1+r_l}/t_l,v_l/t_l) \nonumber \\
 & & ~~~~~~~~+\frac{4K_l^2t_l}{(1+r_l-v_l)^2}~f_1(v_l/t_l) \ ,
     \label{bfrge3} \\
 & & \frac{dU_l}{dl}=-24K_l^2\frac{\sqrt{1+r_l}}{(1+r_l-v_l^2)^2}
     ~g_3(\sqrt{1+r_l}/t_l,v_l/t_l) \nonumber \\
 & & ~~~~~~~~+24K_l^2\frac{v_l}{(1+r_l-v_l^2)^2}
     ~g_4(\sqrt{1+r_l}/t_l,v_l/t_l) \nonumber \\
 & & ~~~~~~~~-\frac{6K_lU_l}{\sqrt{1+r_l}(1+r_l-v_l^2)}~
     g_2(\sqrt{1+r_l}/t_l,v_l/t_l) \nonumber \\
 & & ~~~~~~~~+\frac{12K_lU_lt_l}{(1+r_l-v_l^2)^2}~f_1(v_l/t_l)
     \nonumber \\
 & & ~~~~~~~~-\frac{3U_l^2}{8(1+r_l)^{3/2}}~f_2(\sqrt{1+r_l}/t_l)
     \ , \label{bfrge4}
\end{eqnarray}
where $t=1/\beta$, $K=(\lambda v)^2/(2\pi^2)$, $U=u/(2\pi^2)$ and
\begin{eqnarray*}
 & & f_1(x)=x\coth{ \! \left(\frac{x}{2}\right)} \ , \\
 & & f_2(x)=\coth{ \! \left(\frac{x}{2}\right)}+\frac{x}{2}
           \sinh^{-2}{ \! \left(\frac{x}{2}\right)} \ , \\
 & & g_1(x,y)=\frac{f_1(x)-f_1(y)}{x^2-y^2} \ , \\
 & & g_2(x,y)=\frac{x^2+y^2}{x^2-y^2}~\coth{ \! \left(\frac{x}{2}\right)}
     +\frac{x}{2}\sinh^{-2}{ \! \left(\frac{x}{2}\right)} \ ,
     \\
 & & g_3(x,y)=\frac{x^2+3y^2}{x^2-y^2}~\coth{ \! \left(\frac{x}{2}\right)}
     +\frac{x}{2}\sinh^{-2}{ \! \left(\frac{x}{2}\right)} \ ,
     \\
 & & g_4(x,y)=\frac{3x^2+y^2}{x^2-y^2}~\coth{ \! \left(\frac{y}{2}\right)}
     -\frac{y}{2}\sinh^{-2}{ \! \left(\frac{y}{2}\right)} \ .
\end{eqnarray*}
Here the rescaled quantities have explicit $l$ dependence
indicated, e.g. $t_l$, while quantities without $l$ dependence
(e.g. $T$) refer to physical quantities.

Since $K>0$, we always have $dv_l/dl<0$ from Eq. (\ref{bfrge5}),
and thus $v_l$ will flow to zero. Therefore, we may set $v_l=0$ in
Eqs. (\ref{bfrge2})
---(\ref{bfrge4}), yielding
\begin{eqnarray}
 & & \frac{dr_l}{dl}=2r_l-\frac{4K_lt_l}{1+r_l}+\frac{(U_l+8K_l)t_l}
     {4(1+r_l)}f_1 \! \! \left(\frac{\sqrt{1+r_l}}{t_l}\right) ,
     ~~~~~~\label{bfrge20} \\
 & & \frac{dK_l}{dl}=\frac{8K_l^2t_l}{(1+r_l)^2}-\frac{K_l}{4}
    \frac{U_l+10K_l}{(1+r_l)^{3/2}}f_2 \! \! \left(\frac{\sqrt{1+r_l}}
    {t_l}\right) , \label{bfrge30} \\
 & & \frac{dU_l}{dl}=\frac{24K_lU_lt_l}{(1+r_l)^2}-\frac{3}{8}
     \frac{(U_l+8K_l)^2}{(1+r_l)^{3/2}}f_2 \! \! \left(
     \frac{\sqrt{1+r_l}}{t_l}\right) . \label{bfrge40}
\end{eqnarray}
The solution of Eq. (\ref{bfrge1}) is simply given by
\begin{equation}
 t_l=te^l \ . \label{rge1sol}
\end{equation}
In the next section, we shall solve Eqs. (\ref{bfrge20}) ---
(\ref{bfrge40}) approximately by neglecting the possible
logarithmic corrections.

\section{Solution of scaling equations}
\label{sol}

Scaling stops when $|r_l|\sim 1$. One must distinguish two
regimes: $t_l\ll 1$ and $t_l\gg 1$. The former corresponds to the
quantum ($T=0$) regime where the quantum fluctuations dominate the
low-energy physics. Otherwise, it is the continuum high $T$ regime
where the thermal fluctuations become important.

\subsection{Quantum regime}

In the quantum regime, the nonlinear dependence of the RG
functions (those terms appearing at the R.H.S. of Eqs.
(\ref{bfrge20}) --- (\ref{bfrge40})) on $r_l$ will not be
important. Thus, we shall consider the limiting equations, valid
for $r, K, U\ll 1$
\begin{eqnarray}
 & & \frac{dr_l}{dl}=2r_l-4K_lt_l+ \! \! \left(\frac{U_l}{4}+2K_l
     \right)t_lf_1 \! \! \left(\frac{1}{t_l}\right) ,
     \label{bfrge21} \\
 & & \frac{dK_l}{dl}=8K_l^2t_l-\frac{K_l}{4}(U_l+10K_l)f_2 \! \!
     \left(\frac{1}{t_l}\right) , \label{bfrge31} \\
 & & \frac{dU_l}{dl}=24K_l(U_l+4K_l)t_l- \! \frac{3}{8}(U_l+8K_l)^2
     f_2 \! \! \left(\frac{1}{t_l}\right) . ~~~~~~\label{bfrge41}
\end{eqnarray}

To obtain the condition on $T$ for the occurrence of the quantum
regime, one may set $t=0$ in Eqs. (\ref{bfrge21}) ---
(\ref{bfrge41}), yielding
\begin{eqnarray}
 & & \frac{dr_l}{dl}=2r_l+\frac{1}{6}{\mathcal X}_l+\frac{1}{12}U_l
     \ , \label{bfrge22} \\
 & & \frac{d{\mathcal X}_l}{dl}=-\frac{3}{8}{\mathcal X}_l^2 \ ,
     \label{bfrge32} \\
 & & {\mathcal X}_l\frac{d{\mathcal Y}_l}{d{\mathcal X}_l}=\frac{1}{9}
     ({\mathcal Y}_l-4)({\mathcal Y}_l-1) \ . \label{bfrge42}
\end{eqnarray}
Here ${\mathcal X}_l=U_l+12K_l$ and ${\mathcal Y}_l=U_l/{\mathcal
X}_l$. The solution of Eqs. (\ref{bfrge32}) and (\ref{bfrge42}) is
written as
\begin{eqnarray}
 & & {\mathcal X}_l=\frac{x_0}{1+3x_0l/8} \ , \nonumber \\
 & & U_l=\frac{d_0{\mathcal X}_l^{1/3}-4}{d_0{\mathcal X}_l^{1/3}-1}
     ~{\mathcal X}_l\ , \label{rgsol11}
\end{eqnarray}
where $x_0=U+12K$ and the value of $d_0$ is chosen such that
$U_l=U$ when ${\mathcal X}_l=x_0$. (In fact,
$d_0x_0^{1/3}=4+U/(4K)$. Therefore, $d_0\gg 1$ for $K,U\ll 1$.)

\begin{figure}
\begin{center}
 \includegraphics[width=0.9\columnwidth]{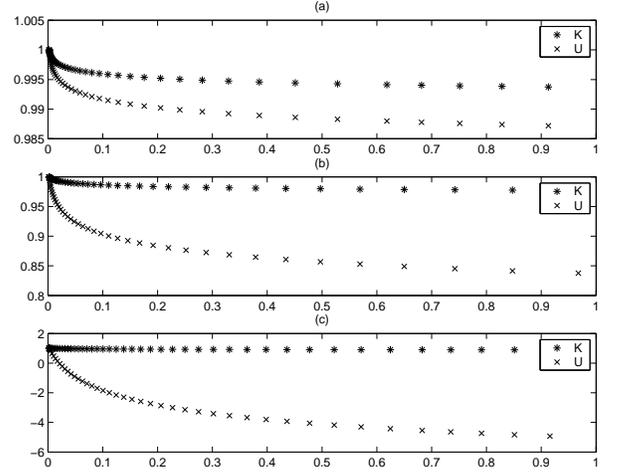}
 \caption{Typical solutions of Eqs. (\ref{bfrge22})---(\ref{bfrge42})
          with the initial conditions (a) $r=0.002$, $K=0.0003$, and $U=0.005$,
          (b) $r=0.002$, $K=0.003$, and $U=0.005$, and (c) $r=0.002$, $K=0.03$,
          and $U=0.005$. The $x$-axis is $r_l$ and the $y-axis$ is $K_l/K$ or
          $U_l/U$.}
 \label{rgfig1a}
\end{center}
\end{figure}

By solving Eqs. (\ref{bfrge22}) --- (\ref{bfrge42}) numerically,
we notice that the value of $K_l$ remains almost unchanged when
$r_l\leq 1$, as shown in Fig. \ref{rgfig1a}. However, the behavior
of $U_l$ depends on the sign of the function $G(r,U,K)$, roughly
given by
\begin{equation}
 G(r,U,K)\approx \frac{16}{3x_0}\left[ \! \left(1+\frac{U}{16K}\right)^3
         \! -1\right] \! -\ln{(1/r)} \ . \label{g1}
\end{equation}
(Equation (\ref{g1}) is determined from Eq. (\ref{rgsol11}) by
requiring that $U_{l^*}=0$ or $d_0{\mathcal X}_{l^*}^{1/3}=4$
where $r_{l^*}=1$. The condition $r_{l^*}=1$ roughly gives rise to
$2l^*\approx \ln{(1/r)}$.) (i) As $G(r,U,K)>0$, $U_l>0$ and its
magnitude does not change much for $r_l\leq 1$, as shown in Fig.
\ref{rgfig1a} (a) and (b). (ii) As $G(r,U,K)<0$, the value of
$U_l$ becomes negative rapidly before $r_l=1$ as shown in Fig.
\ref{rgfig1a} (c).

For given $K$ and $U$, we may increase the value of $r$ starting
with $r=0$. In the beginning, $G(r,K,U)<0$ and $U$ becomes
negative when scaling stops. Thus, a $\phi^6$ term must be
included in the effective action. A mean-field theory in this case
gives rise to $\langle \phi \rangle \neq 0$. This corresponds to
the ASF phase. On the other hand, for sufficiently large $r$ such
that $G(r,U,K)>0$, $U$ is still positive when scaling stops, and
thus a mean-field theory gives rise to $\langle \phi \rangle =0$.
This corresponds to the MSF phase. Consequently, there are two
phases at zero temperature separated by a transition point,
denoted by $r_{c0}$. The value of $r_{c0}$ can be estimated by
requiring that $G(r_{c0},U,K)=0$. A negative value of $U$ is
usually interpreted as a fluctuation-induced first-order phase
transition.\cite{AZ} Therefore, the QPT at $T=0$ is of first
order. In the following, we shall focus on case (i), i.e. the
Z$_2$ symmetric (MSF) phase.

In case (i), because the values of $K_l$ and $U_l$ do not change
significantly when scaling stops, we may set $K_l={\mathcal K}>0$
and $U_l={\mathcal U}>0$ in Eq. (\ref{bfrge22}) for simplicity,
where ${\mathcal K}/K\approx 1$ and ${\mathcal U}/U\leq 1$. Within
this approximation, the logarithmic corrections are neglected.
(The approximation we use amounts to neglecting the $l$ dependence
of ${\mathcal X}_l$. Inclusion of it will result in logarithmic
corrections.) Setting $r_l=1$, solving for $l$, substituting the
result into Eq. (\ref{rge1sol}), and demanding $t_l\ll 1$, one may
obtain the condition for the occurrence of the quantum regime in
the MSF phase:
\begin{equation}
 t\ll \left(\frac{r+A_1}{1+A_1}\right)^{1/2} \ , \label{cross1}
\end{equation}
where $A_1={\mathcal K}+{\mathcal U}/8$.

In terms of Eq. (\ref{lagc2}), the crossover line between the
quantum and continuum high $T$ regimes in the MSF phase,
corresponding to $r>r_{c0}$, can be expressed by
\begin{equation}
 \left(\frac{T}{c_0}\right)^2\approx t_0+\frac{c_0\Lambda^2}{2\pi^2}
      \left(u_2\lambda_0^2+\frac{u_0}{8}\right) \ , \label{cross2}
\end{equation}
as shown in Fig. \ref{phase}. Here the values of $c_0$, $t_0$,
$\lambda_0$, and $u_0$ near the transition point is given by Eq.
(\ref{lagc2}). In addition, the momentum UV cutoff $\Lambda$ can
be estimated by $\Lambda \approx
min\{2^{5/4}\sqrt{mg_0}~n^{1/4},\sqrt{u_2M/2}~n^{1/2}\}$, where
the former and the latter arise from the gap of the real part of
$\Psi_a$ and that of the amplitude fluctuations of $\Phi_m$,
respectively. To obtain Eq. (\ref{cross2}), we have neglected
$1+A_1$ in Eq. (\ref{cross1}) because $A_1\ll 1$. Further,
${\mathcal K}$ and ${\mathcal U}$ are replaced by $K$ and $U$,
respectively.

Finally, we determine the correlation length in the quantum
regime. To do so, we may set $K_l={\mathcal K}$ and $U_l={\mathcal
U}$ in Eq. (\ref{bfrge21}) and performing low-temperature
expansion, yielding
\begin{equation}
 \frac{dr_l}{dl}=2r_l+2A_1-4{\mathcal K}te^l \ . \label{bfrge24}
\end{equation}
The solution of Eq. (\ref{bfrge24}) is written as
\begin{equation}
 r_l=e^{2l}\left[r+A_1\left(1-e^{-2l}\right)-4{\mathcal K}t
    \left(1-e^{-l}\right)\right] . \label{bsol1}
\end{equation}
The correlation length is given by $\xi \sim e^{l^*}$ where
$r_{l^*}=1$. From Eq. (\ref{bsol1}), one may obtain
\begin{equation}
 \xi \sim (r+A_1-A_2t)^{-1/2} \ , \label{xi2}
\end{equation}
with $A_2=4{\mathcal K}$. Therefore, $\xi$ is always finite for
given $t$ satisfying the inequality (\ref{cross1}), as expected
for a first-order phase transition. To sum up, the phase
transition between the MSF and ASF phases in the quantum (low
temperature) regime is of weakly first-order.

\subsection{Continuum high $T$ regime}

In this regime, it is convenient to divide the scaling into two
steps: $t_l\ll 1$ and $t_l\gg 1$. This introduces multiplicative
errors of order unit coming from the imprecise treatment of the
crossover regime $t_l\sim 1$. In the first step, we integrate over
$l$ from $0$ to $\tilde{l}$, such that $t_{\tilde{l}}=1$. Next we
consider the case $t_l\gg 1$. In this case, it is more convenient
to define the coupling $V_l=t_lU_l$. We would like to show that
the RG equations for $r_l$ and $V_l$ in the continuum high $T$
regime are identical to those for the $3d$ Ising model.

We start with Eqs. (\ref{bfrge20}) and (\ref{bfrge40}) and take
$t_l\gg 1$. Using $f_1(0)=2$ and $f_2(x\rightarrow 0)\approx 4/x$,
we obtain
\begin{eqnarray}
 & & \frac{dr_l}{dl}=2r_l+\frac{V_l}{2(1+r_l)} \ , \label{bfrge23}
     \\
 & & \frac{dV_l}{dl}=V_l-\frac{3V_l^2}{2(1+r_l)^2} \ . \label{bfrge43}
\end{eqnarray}
Equations (\ref{bfrge23}) and (\ref{bfrge43}) are identical to the
one-loop RG equations for the $3d$ Ising model. Therefore, we
conclude that the transition between the MSF and ASF phases in the
continuum high $T$ regime is of second order and belongs to the
$3d$ Ising universality class. Near the transition point, the
solution of Eqs. (\ref{bfrge23}) and (\ref{bfrge43}) is given by
\begin{equation}
 r_l-r^*\approx (\tilde{r}-r^*)e^{l/\nu_3} \ , ~~
 V_l\approx V^* \ , \label{sol2}
\end{equation}
where $\nu_3\approx 0.63$ is the correlation length exponent for
the $3d$ Ising model and $\tilde{r}=r_{\tilde{l}}$, which are
determined by the RG equations in the quantum regime. Here $r^*$
and $V^*$ are the fixed points of Eqs. (\ref{bfrge23}) and
(\ref{bfrge43}), which within the $\epsilon$ expansion corresponds
to the Wilson-Fisher fixed point, where $\epsilon =4-d$ and $d$ is
the spatial dimensions. The exact values of $r^*$ and $V^*$ are
nonuniversal, and the determination of them is beyond the present
field-theoretical approach.

To determine $\tilde{r}$, we shall focus on the MSF phase. In this
case, one may neglect the possible logarithmic corrections and set
$K_l=\overline{{\mathcal K}}>0$ and $U_l=\overline{{\mathcal
U}}>0$ where $\overline{{\mathcal K}}/K\approx 1$ and
$\overline{{\mathcal U}}/U\leq 1$. (Note that both
$\overline{{\mathcal K}}$ and $\overline{{\mathcal U}}$ may not be
equal to ${\mathcal K}$ and ${\mathcal U}$.) Then, from Eq.
(\ref{bfrge21}), we obtain
\begin{equation}
 \tilde{r}=\frac{r+B_1}{t^2}-\frac{B_2}{t}+C \ , \label{rgini1}
\end{equation}
where $B_1=\overline{{\mathcal K}}+\overline{{\mathcal U}}/8$,
$B_2=4\overline{{\mathcal K}}$, and
\begin{eqnarray*}
 C=3\overline{{\mathcal K}}-\frac{\overline{{\mathcal U}}}{8}
  +4B_1\int^1_0 \! \! dy~y^{-3}n_B(y^{-1}) \ ,
\end{eqnarray*}
with $n_B(x)=(e^x-1)^{-1}$.

Scaling stops when $|r_{l^*}-r^*|=1$. The correlation length is
given by $\xi \sim e^{\tilde{l}}e^{l^*}=t^{-1}e^{l^*}$. Equation
(\ref{sol2}) gives rise to
\begin{eqnarray*}
 l^*\approx -\nu_3\ln{ \! \left|\frac{r+B_1}{t^2}-\frac{B_2}{t}+B_3\right|}
    \ ,
\end{eqnarray*}
where $B_3=C-r^*$ is a nonuniversal constant. As a result, we get
\begin{equation}
 \xi \sim t^{-1}\left[\frac{|r-r_c(t)|}{t^2}\right]^{-\nu_3} \ ,
     \label{xi1}
\end{equation}
where $r_c(t)=-B_1+B_2t-B_3t^2$ denotes the transition point
between the MSF and ASF phases for given temperature. Equation
(\ref{xi1}) is valid up to logarithmic corrections.

Finally, we consider the RG equation for $K_l$ in the continuum
high $T$ regime, which can be written as
\begin{equation}
 U_l\frac{dR_l}{dU_l}=\frac{4}{3}~R_l^2-\frac{1}{3}~R_l \ ,
     \label{bfrge33}
\end{equation}
where $R=K/U$. Solving Eq. (\ref{bfrge33}) gives rise to
\begin{equation}
 K_l\approx \frac{V^*}{4}~e^{-l} \ , \label{sol3}
\end{equation}
near the transition line. When scaling stops, we may perform the
perturbation theory in $K_l$ and $V_l$ as long as $V^*,
K_{l^*}=V^*/(4\xi t)\ll 1$. Within the spirit of the $\epsilon$
expansion, it is indeed the case because $V^*=O(\epsilon)$.

\section{Damping rate of collective excitations in the MSF phase}
\label{damp}

The damping in the collective excitations of condensates will
become much more severe near the second-order transition line due
to the coupling to the critical fluctuations. Therefore, a
measurement of the damping in the collective modes may serve as a
signature of the second-order Ising transition between the MSF and
ASF phases. In this section, we shall use the RG equations to
calculate the damping rate of the collective mode in the MSF phase
near the second-order transition line, arising from the coupling
to the critical fluctuations. Since the identification of the
Ising transition is established within the $\epsilon$ expansion,
the following RG-improved perturbative calculation is supposed to
be understood within such a context.

The full propagator of the $\theta$ field can be written as
$D^{-1}(i\omega_n,\bm{k})=D^{-1}_0(i\omega_n,\bm{k})+\Sigma_{\theta}(i\omega_n,\bm{k})$
where $\omega_n=2n\pi /\beta$, $\Sigma_{\theta}$ is the
self-energy of the $\theta$ field, and
\begin{eqnarray*}
 D_0(k)=\frac{1}{\omega_n^2/v^2+\bm{k}^2} \ ,
\end{eqnarray*}
is the free propagator of the $\theta$ field. The spectrum is
determined by
\begin{eqnarray*}
 D^{-1}(i\omega_n\rightarrow \omega +i0^+,\bm{k})=0 \ ,
\end{eqnarray*}
which gives rise to $\omega =\epsilon_{\bm{k}}-i\gamma_{\bm{k}}$
where $\gamma_{\bm{k}}$ denotes the damping rate and
$\epsilon_{\bm{k}}$ is the dispersion relation of the collection
mode. In the long wavelength limit,
$\epsilon_{\bm{k}}=\overline{v}k$ where $k=|\bm{k}|$ and
$\overline{v}$ is the dimensionless renormalized velocity measured
in $c$. By solving the equation, one may obtain
\begin{equation}
 \gamma_{\bm{k}}=-\frac{v^2}{2\epsilon_{\bm{k}}}~{\mathrm Im}
        \Sigma_{\theta}(\omega+i0^+,\bm{k}) \ , \label{bdamp1}
\end{equation}
with the understanding that $\omega$ is replace by
$\epsilon_{\bm{k}}$. Near the second-order transition line,
$\gamma_{\bm{k}}=F(\bm{k},K,U,t,\Delta)$ where $\Delta =r-r^*$ and
we have
\begin{eqnarray*}
 F(\bm{k},K,U,t,\Delta)=e^{-l}F(e^l\bm{k},K_l,U_l,e^lt,\Delta_l) \ ,
\end{eqnarray*}
because the scaling dimension of $\gamma_{\bm{k}}$ is
$[\gamma_{\bm{k}}]=1$. Scaling stops at $\Delta_{l^*}=1$, and we
obtain
\begin{equation}
 \gamma_{\bm{k}}=\xi^{-1}\Phi (\xi \bm{k},K_{l^*},U_{l^*},\xi t)
       \ , \label{bdamp2}
\end{equation}
where $\Phi (x_1,x_2,x_3,x_4)=F(x_1,x_2,x_3,x_4,1)$ is the scaling
function.

To the one-loop order, the scaling function $\Phi$ can be written
as
\begin{eqnarray}
 & & \Phi (\xi \bm{k},K_{l^*},U_{l^*},\xi t) \nonumber \\
 & & =-\frac{K_{l^*}\epsilon_{\xi \bm{k}}}{8}\int \! \! d^3p~\frac{
     \delta [\epsilon_{\xi \bm{k}}-E(\bm{p})+E(\xi \bm{k}+\bm{p})]}
     {E(\bm{p})E(\xi \bm{k}+\bm{p})} \nonumber \\
 & & \times \left\{n_B \! \left[\frac{E(\bm{p})}{\xi t}\right]
     -n_B \! \left[\frac{E(\xi \bm{k}+\bm{p})}{\xi t}\right]
     \right\} \nonumber \\
 & & +\frac{K_{l^*}\epsilon_{\xi \bm{k}}}{16}\int \! \! d^3p~\frac{
     \delta [\epsilon_{\xi \bm{k}}-E(\bm{p})-E(\xi \bm{k}+\bm{p})]}
     {E(\bm{p})E(\xi \bm{k}+\bm{p})} \nonumber \\
 & & \times \left\{n_B \! \left[\frac{E(\bm{p})}{\xi t}\right]
     -n_B \! \left[-\frac{E(\xi \bm{k}+\bm{p})}{\xi t}\right]
     \right\} \nonumber \\
 & & -\frac{K_{l^*}\epsilon_{\xi \bm{k}}}{16}\int \! \! d^3p~\frac{
     \delta [\epsilon_{\xi \bm{k}}+E(\bm{p})+E(\xi \bm{k}+\bm{p})]}
     {E(\bm{p})E(\xi \bm{k}+\bm{p})} \nonumber \\
 & & \times \left\{n_B \! \left[\frac{E(\bm{p})}{\xi t}\right]
     -n_B \! \left[-\frac{E(\xi \bm{k}+\bm{p})}{\xi t}\right]
     \right\}. ~~~~~~\label{bdamp3}
\end{eqnarray}
Here $E(\bm{p})=\sqrt{\bm{p}^2+1}$ and $n_B(x)=(e^x-1)^{-1}$. For
small $k$, we find that the damping rate to the lowest order in
$k$ is given by\cite{V}
\begin{equation}
 \gamma_{\bm{k}}=\frac{\pi V^*\overline{v}^2}{16}\left(\frac{k}{\xi t}
       \right)\left[\exp{ \! \left(\frac{1}{\sqrt{1-\overline{v}^2}\xi t}\right)}
       \! \! -1\right]^{-1} , \label{bdamp4}
\end{equation}
where from Eq. (\ref{xi1}) $\xi t\sim
\left[|r-r_c(t)|/t^2\right]^{-\nu_3}$. From Eq. (\ref{bdamp4}), we
see that away from the critical region, the damping of the
collective excitations arising from the coupling to the critical
fluctuations is exponentially small. On the other hand, $\xi t$
diverges near the transition line, and thus within the critical
region, $\gamma_{\bm{k}}$ can be approximated as
\begin{equation}
 \gamma_{\bm{k}}=\frac{\pi V^*}{16}\overline{v}^2\sqrt{1-\overline{v}^2}
        k+O \! \left(\frac{1}{\xi t}\right) . \label{bdamp5}
\end{equation}
Therefore, the collective mode is heavily damped within the
critical region such that no well-defined collective excitations
exist there on account of
$\gamma_{\bm{k}}/\epsilon_{\bm{k}}=O(1)$.

Due to the coupling to the amplitude fluctuations, the collective
mode in the BEC already acquires a damping rate which exhibits
dramatic temperature dependence as given by Ref. \onlinecite{LIU}.
For example, in the low temperature limit, it is of the form:
$\gamma_{\bm{k}}\propto T^4k$.\cite{LIU} Compared with that
result, the damping rate due to the coupling to the critical
fluctuations in the critical region is indeed much stronger than
the one arising from the coupling to the amplitude fluctuations.
However, a careful measurement of the damping rate near the
second-order transition line must be conducted to extract the $3d$
Ising correlation length exponent $\nu_3$. Because within the
critical region the damping is insensitive to the variation of
temperature as shown by Eq. (\ref{bdamp5}).

\section{Conclusion}
\label{con}

We study the QPT between the MSF and ASF phases in terms of RG.
Our calculations suggest that in some parameter region this
transition is of first order as long as the temperature is below
the transition temperature from the normal Bose gas to the
superfluid phase. However, in the other parameter region, the
physics of the QPT is more interesting. It is of weakly
first-order in the low temperature regime where the inequality
given by Eq. (\ref{cross1}) is satisfied. As the temperature is
raised to the continuum high $T$ regime where the inequality given
by Eq. (\ref{cross1}) is reversed, the transition becomes a
second-order one, belonging to the $3d$ Ising universality class.
This second-order transition in the continuum high $T$ regime is
guaranteed by two facts: (i) Those terms containing the coupling
constant $K$ in the one-loop RG equations exactly cancel each
other in the high temperature limit. (ii) The coupling constant
$K$ becomes an irrelevant operator around the $3d$ Ising
(Wilson-Fisher) fixed point in the continuum high $T$ regime and
flows to zero. A schematic phase diagram is shown in Fig.
\ref{phase}, and the crossover line between the quantum and
continuum high $T$ regimes is expressed by the parameters which
are directly related to experimentally measurable quantities, as
given by Eq. (\ref{cross2}). We must emphasized that all our
results are valid only at the temperature much lower than the gap
of amplitude fluctuations in the molecular BEC.

We also calculate the damping in the collective excitation in the
MSF phase near the second-order transition line. Because of the
coupling to the critical fluctuations, the damping will be
strongly enhanced, which may serve as an experimental signature of
the second-order transition. Our calculation shows that the
damping rate of the collective mode in the MSF phase is
insensitive to the variation of temperature near the second-order
transition line, a behavior very different from the one in the
ordinary BEC where the damping is much weaker and exhibits
dramatic temperature dependence.

\acknowledgments

The work of Y.L. Lee is supported by the National Science Council
of Taiwan under grant NSC 93-2112-M-018-009. The work of Y.-W. Lee
is supported by the National Science Council of Taiwan under grant
NSC 93-2112-M-029-007.



\begin{thebibliography}{99}
\bibitem{V} From the RG equation for $v$ [Eq. (\ref{bfrge5})], it
             always flows to zero. Thus, we expect that $\overline{v}<1$.
\end{thebibliography}
\end{document}